\documentclass[showkeys]{revtex4-2}

\usepackage{graphicx}
\usepackage{bm}
\usepackage{amsmath,cleveref}
\usepackage{amssymb}
\usepackage{dcolumn}
\usepackage{xcolor}
\usepackage{titlesec}

\usepackage[latin1]{inputenc}
\usepackage[english]{babel}

\newcommand{\bk}[1]{\left \langle #1 \right \rangle}
\newcommand{\abs}[1]{\left \lvert #1 \right \rvert}

\newcommand{\C}{C^2_N}

\titleformat{\subsubsection}
   {\normalfont\fontsize{9pt}{11pt}\selectfont\bfseries}% apparence commune au titre et au numéro
   {\thesubsubsection.}% apparence du numéro
   {1em}% espacement numéro/texte
   {\centering}% apparence du titre

\begin{document}

%\linenumbers

\title{Accuracy Tests of the Envelope Theory}

\author{Lorenzo \surname{Cimino}}
\email[E-mail: ]{lorenzo.cimino@umons.ac.be}
\thanks{ORCiD: 0000-0002-6286-0722}

\author{Cyrille \surname{Chevalier}}
\email[E-mail: ]{cyrille.chevalier@umons.ac.be}
\thanks{ORCiD: 0000-0002-4509-4309}

\author{Ethan \surname{Carlier}}
\email[E-mail: ]{ethan.carlier@student.umons.ac.be}

\author{Joachim \surname{Viseur}}
\email[E-mail: ]{joachim.viseur@student.umons.ac.be}

\affiliation{Service de Physique Nucl\'{e}aire et Subnucl\'{e}aire,
Universit\'{e} de Mons,
UMONS Research Institute for Complex Systems,
Place du Parc 20, 7000 Mons, Belgium}
\date{\today}

\begin{abstract}
\textbf{Abstract} The envelope theory is an easy-to-use approximation method to obtain eigensolutions for some quantum many-body systems, in particular in the domain of hadronic physics. Even if the solutions are reliable and an improvement procedure exists, the method can lack accuracy for some systems. In a previous work, two hypotheses were proposed to explain the low precision: the presence of a divergence in the potential or the lack of a variational character for peculiar interactions. In the present work, different systems are studied to test these hypotheses. These tests show that the presence of a divergence does indeed cause less accurate results, while the lack of a variational character reduces the impact of the improvement procedure.
%\pacs{03.65.Ge}
%03.65.Ge Solutions of wave equations: bound states
\end{abstract}

\maketitle

\section{Introduction}
\label{sec:intro}

The resolution of equations for many-body quantum systems is always a challenging task. There are numerous approximate computational methods available, each with its own strengths and weaknesses. Among these methods is the envelope theory (ET) \cite{hall80,hall83,hall04}, also known as the auxiliary field method \cite{silv10,silv12}, which stands out as an easy-to-use approach with several advantages: (i) it can handle quite general Hamiltonians with non-standard kinematics in $D$ dimensions; (ii) its computational cost is independent of the number of particles; (iii) in favourable cases, the approximate eigenvalues obtained using the ET can prove to be either lower or upper bounds.

The fundamental idea behind this method is to replace the Hamiltonian $H$ being studied with an auxiliary Hamiltonian $\tilde{H}$ that is solvable \cite{silv12}. In practice, $\tilde{H}$ is chosen to be a many-body harmonic oscillator $H_\text{ho}$, which is exactly solvable for arbitrary dimensions \cite{hall79,cint21}.
Subsequently, an improvement procedure for the method has been developed based on a modification of the global harmonic oscillator quantum number \cite{sema15b}.

The ET can produce quite accurate results for the baryon spectra in the framework of potential models with QCD-inspired interactions \cite{sema15a}. The method has also been used for hybrid mesons in \cite{sema09}. The ET proves particularly useful in situations where the number of particles can be arbitrary large, such as in the large-$N_c$ formulation of QCD \cite{sema07a,sema07b,sema08,buis11,buis12,buis22} (in \cite{sema09,sema07a,sema07b,sema08,buis11,buis12}, a first version of the ET was used without being named as such). The method can also be used in other domains. It has been employed to explore a possible quasi Kepler's third law for quantum many-body systems \cite{sema19a,sema21}. It can be useful for the study of excitons \cite{sema22}. Lastly, the method can be used to validate precise numerical calculations, as in \cite{char15}.

While the method is generally reliable, its accuracy cannot be predicted beforehand. Previous tests \cite{sema15a,sema19,sema20} have demonstrated that the ET yields low relative errors for certain favourable systems, such as those with linear or Gaussian potentials. However, it exhibits lower accuracy for systems with a Coulomb potential. Nevertheless, the application of the improvement procedure enables the recovery of reasonably good results \cite{sema15b}. In a recent investigation of atomic spectra, both the ET and its improved version failed to reproduce binding energies accurately \cite{chev22}. Two hypotheses were suggested to explain this limitation: the presence of a singularity in the potential and the mix of variational character.

The objective of this paper is to confirm these hypotheses by identifying qualitative features that can predict the accuracy of the ET, thereby enhancing its appeal. In Sec. \ref{sec:ET}, the basics of the ET are presented, with a focus on practical aspects. Likewise, Sec. \ref{sec:improved ET} provides a brief outline of the improved ET, highlighting the practical requirements once again. Sec. \ref{sec:test} is entirely dedicated to accuracy tests and their interpretations. The first three subsections examine the impact of divergences on the ET, while the remaining three subsections focus on the role of the variational character. Finally, Sec. \ref{sec:conclu} summarises the initial objectives and draws conclusions regarding the aforementioned qualitative features.

\section{Envelope theory}
\label{sec:ET}

The ET is an approximation method used to determine the spectrum of generic $N$-body Hamiltonians. Consider a system of $N$ identical particles governed by the following Hamiltonian,
\begin{equation}\label{trueH}
    H=\sum_{i=1}^N T(p_i) + \sum_{i<j=2}^N V(r_{ij}),
\end{equation}
where $p_i = \abs{\bm{p_i}}$ and $r_{ij} = \abs{\bm{r}_i-\bm{r}_j}$. The functions $T(x)$ and $V(x)$ are respectively interpreted as a kinetic energy and a two-body potential. Although these functions are general, they must satisfy a few constraints that are not overly restrictive and are aligned with the physical interpretation of $T$ and $V$ (e.g. differentiablility and positiveness of $T$) \cite{sema18b}. The ET provides a good approximation for the spectrum of this Hamiltonian by solving the following set of three equations \cite{sema13},
\begin{subequations}\label{compacteq}
\begin{align}
    &E = N\,T(p_0)+C^2_N\,V(\rho_0),\label{compacteq1}\\
    &N\,T'(p_0)\,p_0=C^2_N\,V'(\rho_0)\,\rho_0,\label{compacteq2}\\
    &Q(N) = \sqrt{C^2_N}\,p_0\,\rho_0.\label{compacteq3}
\end{align}
\end{subequations}
Here, $f'$ denotes the derivative of $f$ with respect to its argument, $C^2_N=N(N-1)/2$ is the number of particle pairs and natural units are used ($c = \hbar = 1$). The demonstration of these equations, commonly referred to as compact equations of the ET, is based on the introduction of the aforementioned auxiliary Hamiltonians. A full description of the proof can be found in \cite{silv12}, while a summary is provided in \cite{sema18c}. The resolution of the compact equations yields an expression of the approximated spectrum $E$ as well as the two parameters $p_0$ and $\rho_0$ in terms of a global quantum number $Q(N)$. The latter is related to the internal radial and orbital quantum numbers $n_\alpha$ and $l_\alpha$ through the following relationship, 
\begin{equation}
    Q(N) = \sum\limits_{\alpha=1}^{N-1} \left(2n_\alpha+l_\alpha+\frac{D}{2}\right),
\end{equation}
where $D$ represents the dimension of the system. The $p_0$ and $\rho_0$ parameters, obtained during the resolution of \eqref{compacteq}, provide an approximation of instructive mean values in the system, 
\begin{align} \label{r0p0}
    &p_0^2 = \bk{p_i^2}, & &\rho_0^2 = \bk{r_{ij}^2}. 
\end{align}
Since all the particles are identical, the $i$ and $j$ subscripts can be indiscriminately chosen in $\{1,...,N\}$. The above mean values are not evaluated on the genuine eigenstates but on the eigenstates of the auxiliary Hamiltonians mentioned in Sec. \ref{sec:intro}. In previous papers, the parameter $r_0^2 = N^2\bk{(\bm{r}_i-\bm{R})^2}$, where $\bm{R}$ denotes the centre of mass position, was used instead of $\rho_0$ to allow for the treatment of one-body potentials. It can be shown that for identical particles both parameters are equivalent up to a multiplicative constant, $r_0^2 = \C \rho_0^2$. Since only two-body potential will be considered in this work, it is preferable to use $\rho_0$.
\vspace{3mm}

For certain favourable Hamiltonians, an approximated energy given by the ET can be interpreted as either an upper or lower bound for the original energy \cite{sema13,hall80,hall83}. To establish this, let us define the following functions,
\begin{align}
& b_T(x^2)=T(x),\\ 
& b_V(x^2)=V(x).
\end{align}
If both the $b_T$ and $b_V$ functions are concave (resp. convex) across their entire domain, then the approximated spectrum given by the ET is an upper (resp. lower) bound. If the second derivative of one of these functions is zero the variational character is determined by the other. If both second derivative vanishes, meaning $H$ is a harmonic oscillator Hamiltonian, the ET provides the exact result. In other cases, no variational character can be guaranteed.
\vspace{3mm}

Before concluding the section, let us mention that generalisations of the ET have been developed. The approximation method can handle Hamiltonians including also one-body potentials and a specific type of $K$-body forces \cite{sema18c}. Furthermore, recent advancements have extended the method to systems composed of $N_a$ identical particles of type $a$ and $N_b$ identical particles of type $b$ \cite{sema20,cimi22}. These generalisations result in a new set of compact equations \eqref{compacteq} while the rest of the discussion remains relevant. 

\section{Improvement of the envelope theory}
\label{sec:improved ET}

While the ET provides reliable approximations, it may lack accuracy for certain Hamiltonians. To address this limitation, a parameter $\phi$ can be introduced into the global quantum number $Q(N)$,
\begin{equation}
Q_\phi(N) = \sum_{\alpha=1}^{N-1} \left(\phi n_\alpha + l_\alpha + \frac{D+\phi-2}{2}\right).
\end{equation}
This addition, inspired by \cite{loba09} where two-body systems are considered, aims to break the strong degeneracy inherent to the harmonic oscillator Hamiltonian. It is important to note that setting $\phi=2$ recovers the original $Q(N)$. The parameter $\phi$ is computed by coupling the ET with a generalisation of the dominantly orbital state method to $N$-body systems \cite{sema15b}. This coupling leads to the following analytical expression for $\phi$,
\begin{align}
\phi=\,&\frac{\lambda}{N\tilde{p}_0 T'(\tilde{p}_0)} \sqrt{\frac{k}{C_N^2\mu}},\label{DOSM_0}\\
&\text{with } \mu=\frac{\tilde{p}_0}{N T'(\tilde{p}_0)},\label{DOSM_1}\\
&\text{with } k=\frac{2 N \tilde{p}_0}{\tilde{\rho}_0^2}\,T'(\tilde{p}_0) + \frac{N \tilde{p}_0^2}{\tilde{\rho}_0^2}\,T''(\tilde{p}_0) + C_N^2 \,V''(\tilde{\rho}_0^2), \label{DOSM_2}\\
&\text{and where }\begin{cases} \label{DOSM_3}
N\,T'(\tilde{p}_0)\,\tilde{p}_0=C^2_N\,V'(\tilde{\rho}_0)\,\tilde{\rho}_0,\\
\sqrt{C^2_N}\,\tilde{p}_0\,\tilde{\rho}_0 =\sum_{\alpha=1}^{N-1}\left(l_\alpha + \frac{D-2}{2}\right).
\end{cases}
\end{align}
The expression presented here is equivalent to the one in \cite{sema15b} but is more concise due to an unnoticed simplification. This formula has been tested for a few physical potentials, concluding that the inclusion of $\phi$ indeed improves the accuracy of the ET. Before proceeding, it is important to note that modifying $Q(N)$ eliminates the guarantee of the variational character, if it exists.

Regarding generalisations, the improvement has also been developed for systems consisting of $N_a$ identical particles plus one different \cite{chev22} and for Hamiltonians that include one-body potentials (only in the case of $N$ identical particles) \cite{sema15b}. In the following, the term ``classical ET" will refer to the ET without the additional parameter $\phi$, in contrast to ``improved ET" which includes it.

\section{Accuracy tests}
\label{sec:test}

In the previous sections, we discussed the ET as an easily applicable approximation method whose accuracy remains nevertheless difficult to predict. While it is known that the improved version of the method enhances its accuracy, quantifying this improvement remains elusive. Recent tests exploring the generalisation of the method for systems of different particles have revealed a significant discrepancy in the reproduction of atomic spectra, even when considering the improvement procedure \cite{sema20,chev22}. In the corresponding articles, two hypotheses were proposed to explain the poor accuracy:
\begin{itemize}
    \item the difficulty in reproducing the divergence of the Coulomb interaction
    \item the mixture of attractive and repulsive potentials, which prevents the method from having a variational character.
\end{itemize}
This article aims to investigate whether these two factors indeed impact the accuracy of the ET through tests of the method. While tests of the ET on physical systems have already been done in \cite{sema15a}, the current work focuses on evaluating the accuracy and the limitations of the method. As a first approach and because we have an accurate numerical code specifically designed for three-body systems at our disposal, we will focus the tests on systems composed of three identical particles with non-relativistic kinematics $T(p) = p^2/2m$. Additionally, we will only consider bosonic ground states (BGS) in three dimensions ($D=3$). For the sake of generality, the formulas will be presented with an arbitrary number of particles $N$, global quantum number $Q_\phi(N)$ and mass $m$. Nonetheless, in the presented graphs, $N$ will be fixed to $3$ and the modified global quantum number for BGS will be reduced to
\begin{equation}
    Q_\phi^{BGS}(N=3) = \phi+1.
\end{equation}
Furthermore, since arbitrary units will be used, $m$ will be set to $1$. Given that the second derivative of $b_T(p)$ vanishes, the variational character is solely determined by the potential. The aforementioned highly accurate numerical method, to which the ET approximations will be compared, is based on an oscillator basis expansion (OBE) \cite{silv00,nunb77}. Results from this numerical method have been abundantly compared to the ones from the literature \cite{basd90}. A sample of these tests is proposed in Table \ref{tab:OBE}. This table also emphasis that OBE is significantly more accurate than the ET. Therefore, results from the OBE will be referred to as ``exact results'' throughout the rest of this article. Nevertheless, let us mention that the numerical cost of OBE is considerably higher than the one of the ET, even for $N=3$.

\begin{table}
    \centering
    \begin{tabular}{rlcrlcrlcrlcrl}
        \multicolumn{2}{c}{$\beta$} &\hspace{2mm}& \multicolumn{2}{c}{Literature} &\hspace{2mm}& \multicolumn{2}{c}{OBE} &\hspace{2mm}& \multicolumn{2}{c}{ET} \\
        \hline\hline
        $-1.$&$\!0 $ && $-0.$&$\!266\,75$ && $-0.$&$\!266\,14$ && $-0.$&$\!125\,00$ \\
        $-0.$&$\!5$ && $-0.$&$\!591\,73$ && $-0.$&$\!591\,68$ && $-0.$&$\!491\,39$ \\
        $0.$&$\!1$ && $1.$&$\!880\,19$ && $1.$&$\!880\,19$ && $1.$&$\!914\,06$ \\
        $0.$&$\!5$ && $2.$&$\!916\,54$ && $2.$&$\!916\,53$ && $3.$&$\!082\,03$ \\
        $1.$&$\!0$ && $3.$&$\!863\,09$ && $3.$&$\!863\,09$ && $4.$&$\!088\,52$ \\
        \hline
    \end{tabular}
    \caption{Ground state energies provided by the ET, the OBE and the literature \cite{basd90} for a system of three identical bosons of unit mass and interacting pairwise with potential $V(r)=0.5 \,\text{sgn}(\beta) r^\beta$.}
    \label{tab:OBE}
\end{table}

\subsection{Negative power potential}

To test the impact of divergence, we consider a negative power-law potential given by $V(r) = -G\,r^{\beta}$, where $\beta < 0$ and $G>0$. This potential becomes more divergent as $|\beta|$ increases. The ET approximation for this potential has previously been computed in \cite{chev22} for a more general Hamiltonian (which includes power-law kinematics). In the case of non-relativistic kinematics, the approximation reduces to

\begin{subequations}
\begin{align}\label{ETNPP}
    &\rho_0 = \left(\frac{N\,Q_\phi(N)^2}{m|\beta|G(\C)^2}\right)^{\frac{1}{2+\beta}},\\
    &E = -(2+\beta)\left[\left(\frac{G}{2}\right)^2(\C)^{2-\beta}\left(\frac{N\,Q_\phi(N)^2}{2m|\beta|}\right)^\beta\right]^{\frac{1}{2+\beta}},
\end{align}
\end{subequations}
where $\phi = \sqrt{2+\beta}$. It has been demonstrated in the same article that this result is an upper bound (as long as $\phi$ is kept to be $2$). When $\beta = -1$, which corresponds to the Coulomb potential, the expected result from \cite{sema15a} is recovered.

\begin{figure}
    \centering
    \includegraphics[scale=0.45]{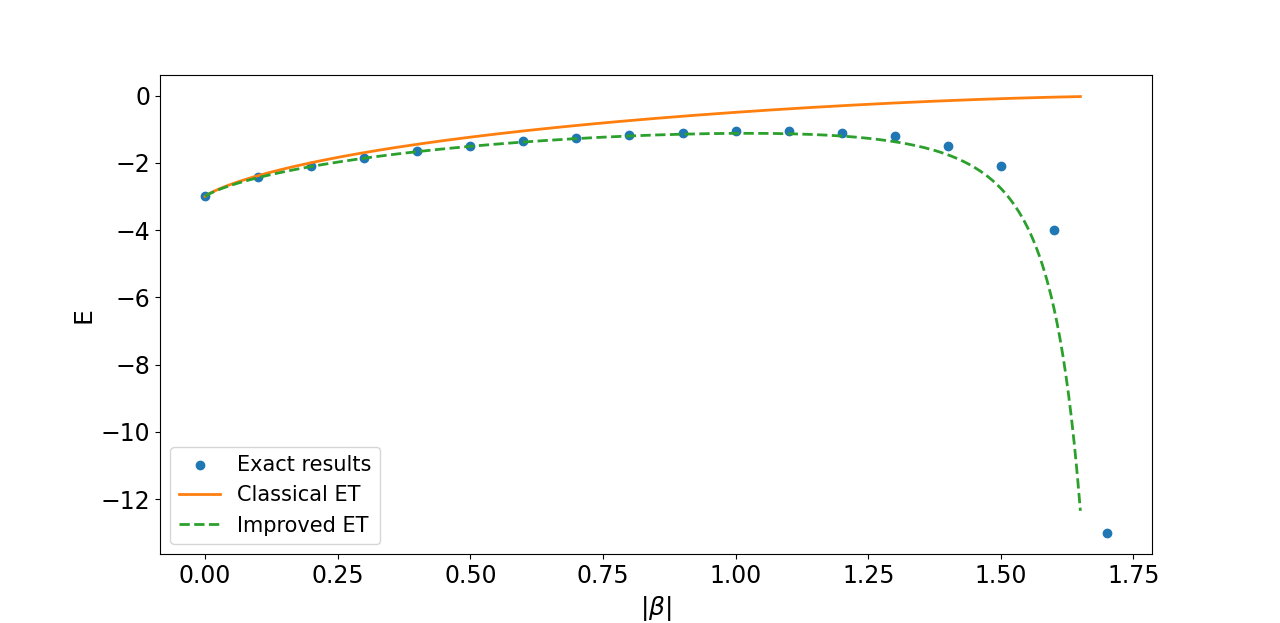}
    \caption{Energy $E$ of a three-body system with $m=1$ and a negative power potential $V(r) = -r^{-\abs{\beta}}$ as a function of $\abs{\beta}$. Exact results (dots), ET results (line) and improved ET results (dashed line). ET and improved ET results have been plotted with a line to easily distinguish them from the exact ones.}
    \label{NPP}
\end{figure}

Comparisons with the exact energies are presented in Fig. \ref{NPP} for $G=1$ and various values of $\beta$. It can be observed that the ET provides reasonably accurate results for small $\abs{\beta}$. However, as $|\beta|$ increases, the relative error also increases. On the other hand, when utilising the improved ET, the behaviour with respect to $\beta$ is well reproduced. For values of $\beta$ close to $-2$, the accuracy of the improved ET results appears to decrease. However, it should be noted that for $\beta<-1$, the results obtained from the oscillator basis expansion should be interpreted with caution because convergence was not totally achieved.

\subsection{Truncated Coulomb potential}

Another way to test the impact of the divergence is to mitigate it through truncation. The Coulomb potential can be modified by introducing a bias distance $d$ that truncates it at the origin: $V(r) = -C/(r+d)$, where $C>0$. An analytical solution of the compact equations \eqref{compacteq2} and \eqref{compacteq3} exists but it requires finding the roots of a cubic equation of the form $t^3 \pm 3t - 2Y = 0$. Let us denote these roots as $F_\pm (Y)$, with their explicit expression given in \cite{silv12, silv09}. Upon inserting $\rho_0$ and $p_0$ into \eqref{compacteq1}, the following approximated energies are obtained,
\begin{subequations}\label{ETTCP}
\begin{align}
    &\rho_0 = d\,y\\
    &E = -\frac{1}{\C}\frac{C}{d}\frac{y+2}{2(y+1)^2}, \\
    &\text{with } y = \frac{\sqrt{A(A+6)}}{3}F_{-}\left(\frac{2A^3+18A^2+27A}{2(A(A+6))^{3/2}}\right) +\frac{A}{3} \text{ and }A=\frac{N}{(C^2_N)^2}\frac{Q_\phi(N)^2}{m\,C\,d}.
\end{align}
\end{subequations}
For $\phi=2$, this energy is demonstrated to be an upper bound of the genuine energy. Although the analytical expression of the spectrum provided by the ET is complicated to write (mainly due to the $F_\pm (Y)$ functions), it simplifies in certain limits. For instance, in the limits $d>>r$ and $d<<r$, \eqref{ETTCP} reduces to
\begin{subequations}
\begin{align}
    &E_{d>>r} = -C^2_N\frac{C}{d} + \frac{3}{2}\left(\frac{N}{C^2_N}\right)^{1/3}\left(\frac{Q_\phi(N)^2C^2}{m\,d^4}\right)^{1/3},\\
    &E_{d<<r} = -\frac{(C^2_N)^3}{N}\frac{mC^2}{2Q_\phi(N)^2} + \frac{(C^2_N)^4}{N^2}\frac{m^2C^3d}{Q_\phi(N)^4}.
\end{align}
\end{subequations}

While an analytical solution for $\rho_0$ and $p_0$ exists, it is also theoretically possible to obtain an analytical expression for $\phi$ by using equations \eqref{DOSM_0} - \eqref{DOSM_3}. However, due to the complexity of the solution mentioned above, the resulting expression for $\phi$  would be too intricate to present here and a numerical evaluation proves to be more comfortable and efficient.
 
\begin{figure}
    \centering
    \includegraphics[scale = 0.45]{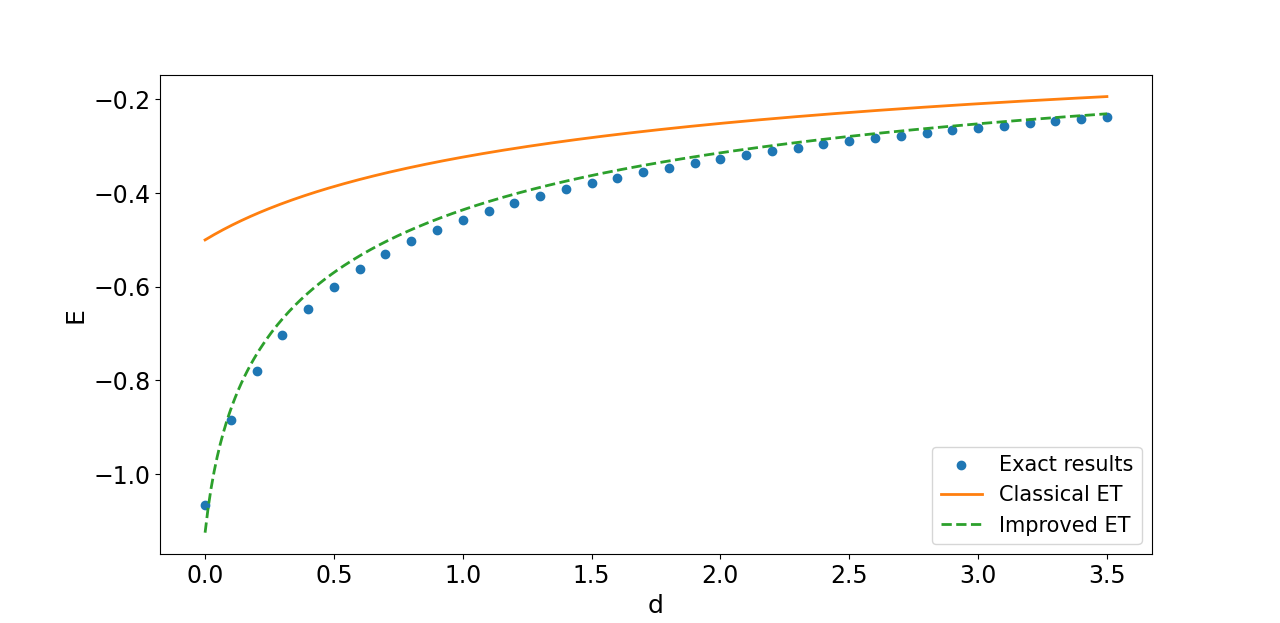}
    \caption{Energy $E$ of a three-body system with $m=1$ and a truncated Coulomb potential $V(r) = -1/(r+d)$ as a function of $d$. Exact results (dots), ET results (line), and improved ET results (dashed line).}
    \label{TCP}
\end{figure}

By varying $d$, it is possible to control the divergence of the potential. When $d=0$, the original Coulomb potential is, of course, recovered. Accuracy tests are presented in Fig. \ref{TCP} for $C=1$. As $d$ increases, the accuracy of the ET improves. Furthermore, for the improved ET, the behaviour is almost exactly reproduced.

\subsection{Exciton potential}

A smoother approach for truncating the Coulomb potential is to replace it with $V(r) = -C/\sqrt{r^2+d^2}$, where $C>0$. This modified potential is commonly used to model electron-hole pairs, also known as excitons \cite{gras17}. By applying equations (\ref{compacteq3}) and (\ref{compacteq2}), the following expression is obtained
\begin{equation}
    \frac{m\,C}{N}\frac{(\C)^2}{Q_\phi(N)^2}\rho_0^4 = (\rho_0^2+d^2)^{3/2}.
\end{equation}
Since the solutions of this quartic equation are complicated to manipulate, numerical computations are preferred. It has been demonstrated that the classical ET provides upper bounds for this potential. In a previous study \cite{sema22}, the ET was employed to investigate this system, but only for two-body systems at $D=1$. As the ET results are obtained through numerical calculations, the same applies to the improved ET.

\begin{figure}
    \centering
    \includegraphics[scale = 0.45]{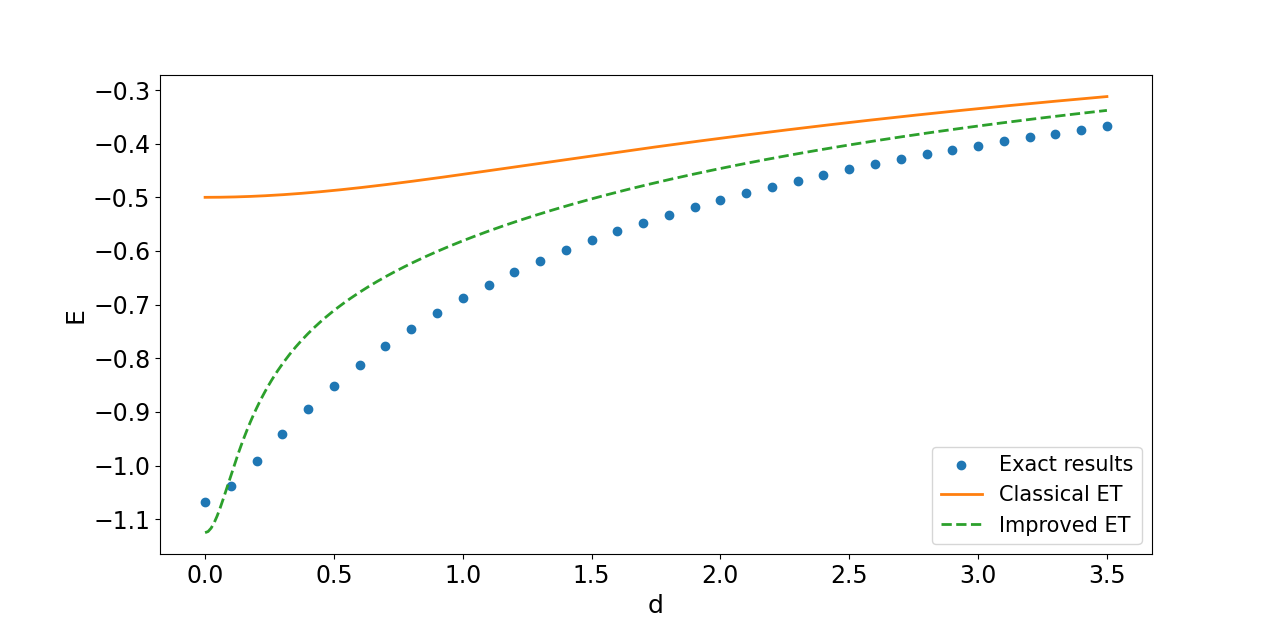}
    \caption{Energy $E$ of a three-body exciton system with $m=1$ and $V(r) = -1/\sqrt{r^2+d^2}$ as a function of $d$. Exact results (dots), ET results (line) and improved ET results (dashed line).}
    \label{EP}
\end{figure}

The conclusions drawn for this potential are similar those of the previous example. As shown in Fig. \ref{EP}, the relative errors decrease as the divergence is smoothed (i.e. as $d$ increases). Concerning the improved results, they better reproduce the dependence of energy on $d$.

\subsection{Cubic plus linear potential}

The second hypothesis proposed to explain the poor accuracy of the ET in the evaluation of atomic spectra was related to its lack of variational character. In atomic systems, the repulsive Coulomb interactions among the electrons tend to provide a lower bound, while the attractive Coulomb interactions with the nucleus tend to provide an upper bound. As a result, predicting the variational character of the ET for atomic systems becomes impossible.

To replicate this situation, a potential consisting of two components, one leading to an upper bound and the other to a lower bound, can be considered. Additionally, to monitor the presence of the variational character, a constant $C$ is introduced. By continuously varying  $C$ from $0$ to $1$, one potential is ``turned on" while the other is ``turned off". Consequently, the transition from $C=0$ to $C=1$ corresponds to a change in the variational character.

 Let us consider a first example to illustrate this approach. On one hand, the cubic potential $V(r) = \alpha \,r^3$ exhibits a lower bound because its function $b_V(r) = \alpha \, r^{3/2}$ is convex. On the other hand, the linear potential $V(r) = \beta \,r$ produces an upper bound since its function $b_V(r) = \beta \,r^{1/2}$ is concave. Therefore, a potential that combines both of these features will preferentially present, depending on the value of $C$, either one variational character or the other,
\begin{equation}
    V(r) = \alpha\, C\,r^3 + \beta\, (1-C)\, r.
\end{equation}
When $C=0$, $V(r)$ reduces to the linear potential and when $C=1$, $V(r)$ reduces to the cubic potential. To obtain the approximate spectrum, the compact equations \eqref{compacteq} must be solved once again. It leads to a quintic equation,
\begin{equation}
    \frac{N}{m}\frac{Q(N)^2}{(\C)^2} - 3\alpha\,C\rho_0^5- \beta(1-C)\rho_0^3 = 0,
\end{equation}
which do not have an easy-to-use analytical solution. Therefore, numerical computations will be performed for both the ET and its improvement. 

\begin{figure}
    \centering
    \includegraphics[scale = 0.45]{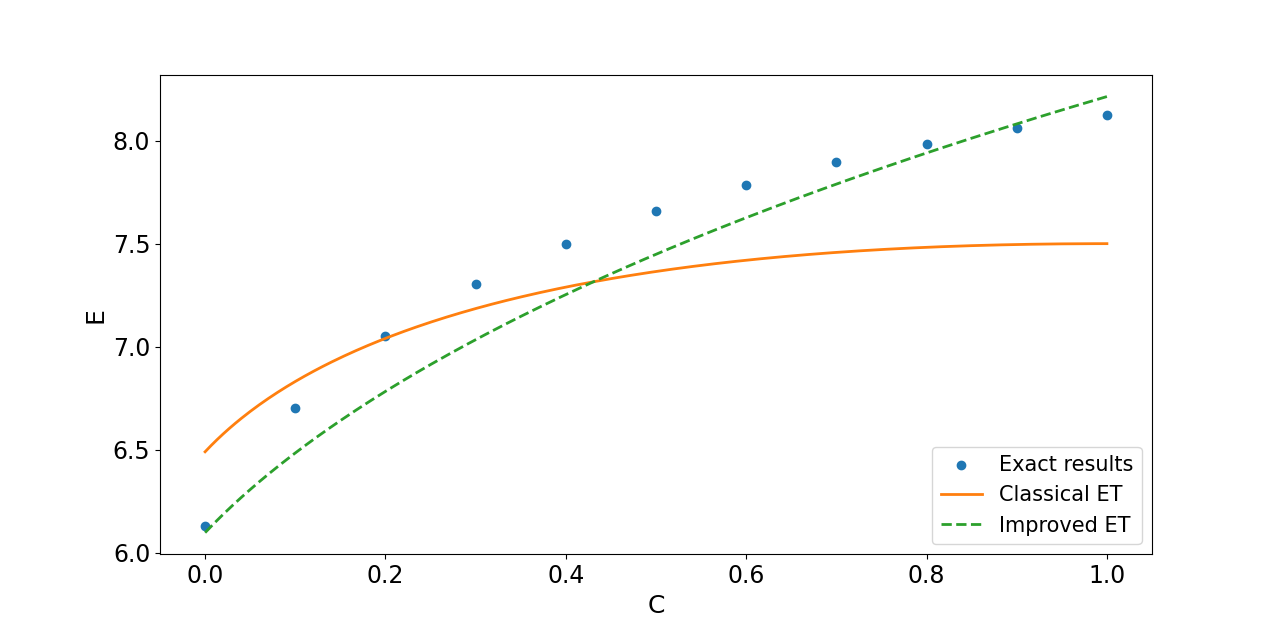}
    \caption{Energy $E$ of a three-body system with $m=1$ and a mixed cubic - linear potential $V(r) = C\,r^3 + (1-C)r$ as a function of $C$. Exact results (dots), ET results (line), and improved ET results (dashed line).}
    \label{MCLiP}
\end{figure}

The results of these computations are presented in Fig. \ref{MCLiP} (with $\alpha$ and $\beta$ set to $1$). As expected, when $C=0$ (pure linear potential), the classical ET provides an upper bound for the energy. On the other hand, when $C=1$ (pure cubic potential), it provides a lower bound. In both cases, the improved ET yields noticeably more accurate results.

For intermediate values of $C$ the situation is different. The errors in the evaluation of the spectrum introduced by the two components of the potential counterbalance each other, resulting in a more precise energy. Since the ET must continuously transition from one variational character to the other, its results even match the exact ones for a specific value of the parameter (in this case, approximately at $C=0.2$). However, due to the unpredictability of the precise location of this crossover point, this feature is impractical to utilise.

Regarding the improved ET, it appears to be less effective in this intermediate range of $C$ values compared to the extreme ones. The variational character can be seen as a safeguard that prevents ET approximations from dropping too low or rising too high. Due to the loss of such a variational guidance, the improved ET encounters challenges in achieving higher accuracy. Regarding the crossing with the exact results, although the feature occurs in this example, having lost the constraint of the variational character, nothing can be predicted in general. In conclusion, it is not guaranteed that the improved ET is consistently more precise than the classical one in such situations. This outcome should be slightly qualified because the dashed curve reproduces the exact shape better and because our analysis focuses solely on the bosonic ground state.

\subsection{Cubic plus logarithmic potential}

To deepen the analysis, a second potential is studied. Considering that a logarithmic interaction $V(r)= \beta \ln(r)$ also produces an upper bound (since the associated $b_V(r)= \beta \ln(\sqrt{r})$ function is concave), the linear contribution in the previous potential can be replaced by a logarithmic one while maintaining a mixture of variational characters,
\begin{equation}
   V(r) = \alpha\,C\,r^3 + \beta(1-C)\ln{(r)}.
\end{equation}
By employing compact equations \eqref{compacteq}, the following quintic equation, which has no trivial solution, is obtained,
\begin{equation}
    \frac{N}{m}\frac{Q(N)^2}{(\C)^2} - 3\alpha\,C\rho_0^5 - \beta(1-C)\rho_0^2 = 0.
\end{equation}
Once again, numerical calculations will be performed for both the ET and its improvement.

\begin{figure}
    \centering
    \includegraphics[scale = 0.45]{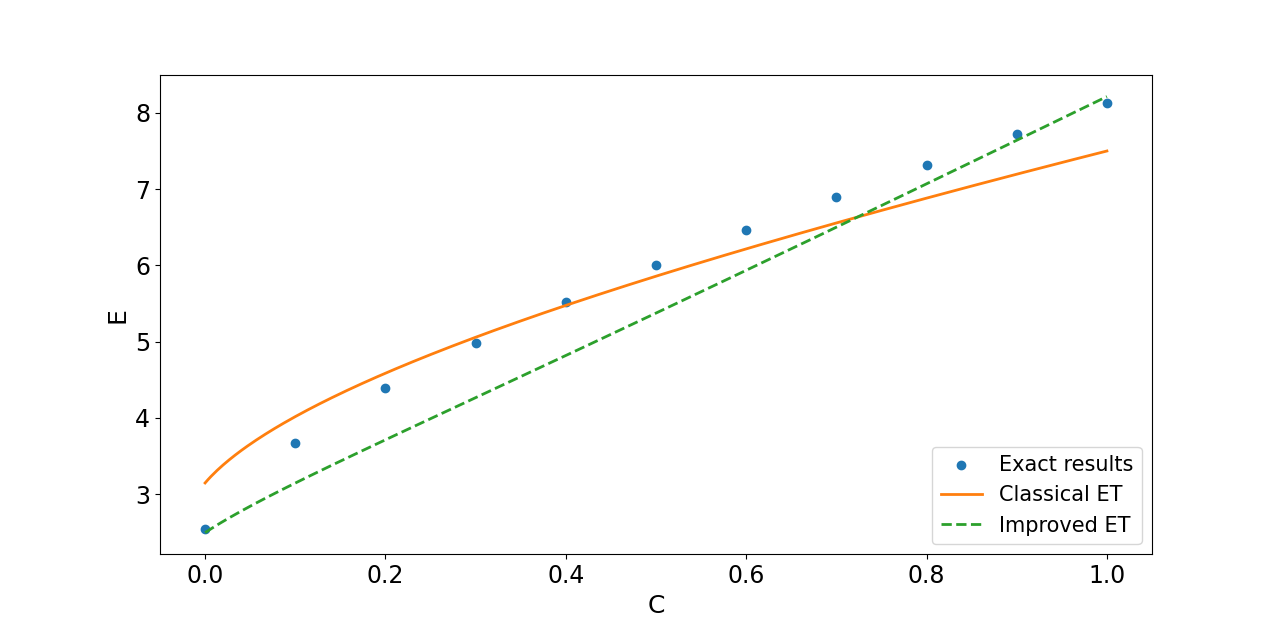}
    \caption{Energy $E$ of a three-body system with $m=1$ and a mixed cubic - logarithmic potential $V(r) = C\,r^3 + (1-C)\ln{r}$ as a function of $C$. Exact results (dots), ET results (line), and improved ET results (dashed line).}
    \label{MCLoP}
\end{figure}

The results of this second test are presented in Fig. \ref{MCLoP} for $\alpha=\beta=1$. They confirm the previous conclusions. When $C\in \{0,1\}$, the classical ET exhibits the expected variational character, and its improvement provides a more accurate estimation. However, for values of $C$ in between these extremes, the upper and lower bounds counterbalance each other, resulting in a more accurate prediction given by the classical ET. As for the improved version of the method, its results are not consistently more accurate than the original ones. In this example, it is not even clear if the shape of the curve is better reproduced by the ET or its improvement.

\subsection{Cubic plus Gaussian potential}

Finally, a last test involving a mixture of variational characters can be performed. Similar to the linear and logarithmic potentials, the Gaussian potential $V(r) = -\alpha e^{-r^2}$ contributes as an upper bound. Therefore, the following combination will be considered,
\begin{equation}
    V(r) = \alpha\,C\,r^3 - \beta(1-C)\,e^{-r^2}
\end{equation}
By solving the compact equations \eqref{compacteq}, the following equation, whose solution cannot be found analytically, is obtained,
\begin{equation}
    \frac{N}{m}\frac{Q(N)^2}{(\C)^2} - 3\alpha\,C\rho_0^5 +2\beta(1-C)e^{-\rho_0^2}\rho_0^4 = 0.
\end{equation}
Numerical computations will still be necessary to evaluate both the ET and its improvement. It is worth noting that the constant $\beta$ must be chosen to be higher than one to ensure the existence of a bound state.

\begin{figure}
    \centering
    \includegraphics[scale = 0.45]{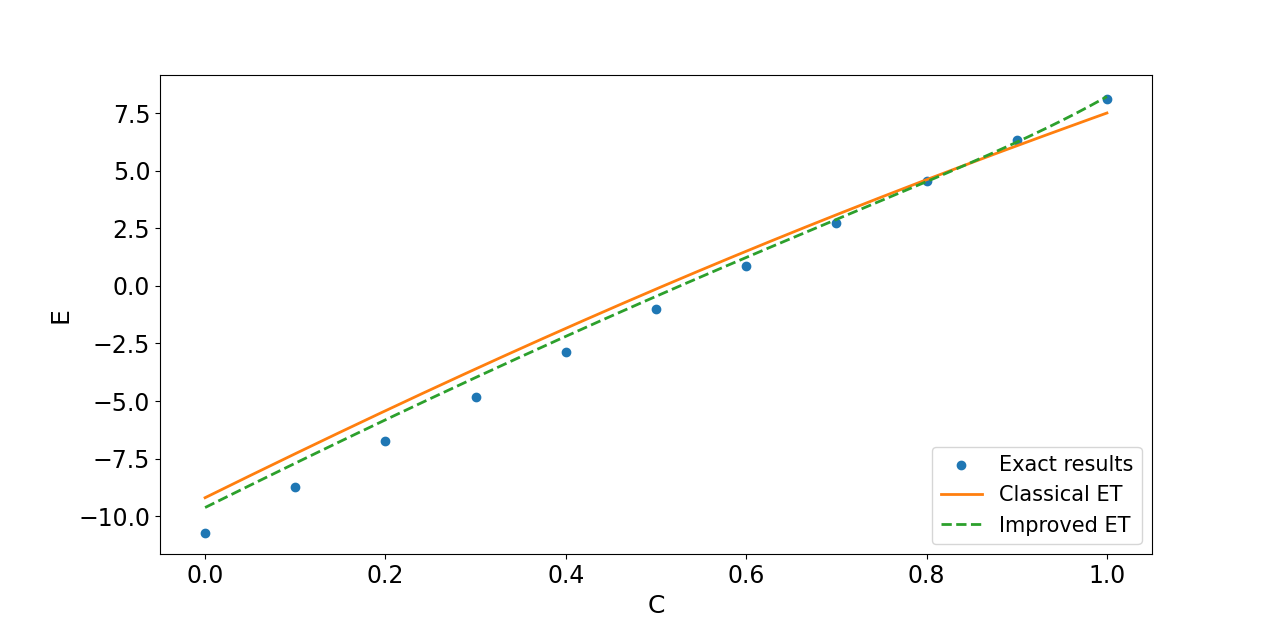}
    \caption{Energy $E$ of a three-body system with $m=1$ and a mixed cubic - Gaussian potential $V(r) = C\,r^3 - 10(1-C)e^{-r^2}$ as a function of $C$. Exact results (dots), ET results (line), and improved ET results (dashed line).}
    \label{MCGP}
\end{figure}

The obtained energies are summarised in Fig. \ref{MCGP} for $\alpha=1$ and $\beta=10$. The situation appears slightly different here. While the results from the classical ET are similar to those obtained in the previous tests, the improved ET provides energies that seem to be consistently more accurate than those obtained with the original method. Both approximation methods intersect the curve of exact results around the same value of $C$. This further emphasises the unpredictability of the behaviour that arises when variationnal character are mixed. 

\section{Concluding remarks}
\label{sec:conclu}

The ET has proven to be a reliable and easily implementable approximation method for $N$ identical particle systems \cite{silv10,silv12,sema13}. However, predicting its performance remains challenging. In certain cases, the ET exhibits a variational character, guiding us toward the exact results. Recently, the ET and its improvement have been extended for systems consisting of $N_a$ identical particles of type $a$ and one particle of type $b$ \cite{sema20,cimi22}. While most tests conducted with this generalisation have yielded conclusive results, the evaluation of atomic spectra has revealed unexpectedly low accuracy, even after improvement \cite{chev22}. To explain this phenomenon, two hypotheses have been proposed: the divergence of the Coulomb interactions, which may be difficult to approximate, and the interplay between attractive and repulsive contributions, hindering the existence of a variational character.

In this study, these two hypotheses have been investigated regarding the bosonic ground state of systems composed of three identical particles. Our findings demonstrate a correlation between the presence of a divergence and decreased accuracy of the ET, while a mixture of variational character can lead to a loss of accuracy in the improved method. These two phenomena, when combined, provide an explanation for the unsatisfactory efficiency of the ET in evaluating atomic spectra. The divergence of Coulomb interactions appears to diminish the accuracy of the original method, whereas the mixture of variational character prevents the improvement from correcting the obtained energies.

In a future research, it would be valuable to confirm these conclusions for systems consisting of two identical particles plus a different one. In such systems, the particle mass ratio could become a significant parameter to explore. Given the significant disparity in mass between electrons and atomic nuclei, investigating this mass dependence could also provide additional insights into the factors affecting the accuracy of the ET in atomic systems. Apart from the issue of the atomic spectra, additional tests on excited states and relativistic kinematics could also further enhance our understanding of the conditions under which the ET provides accurate results. 

\begin{acknowledgments}
L.C. and C.C would thank the Fonds de la Recherche Scientifique - FNRS for the financial support. E.C. and J.V. would thank the University of Mons for their Initiation Research Grant. This work was also supported under Grant Number 4.45.10.08. The authors thank Claude Semay for the careful reading of the manuscript and fruitful discussions.
\end{acknowledgments} 

\section*{Data Availability Statement}
No data associated in the manuscript.

\end{document}